\documentclass[preprint,12pt]{elsarticle}

\usepackage{graphicx}
\usepackage{amssymb}

\usepackage{hyperref}

\journal{Journal of Nuclear Materials}

\begin{document}

\begin{frontmatter}


\title{F-TRIDYN: A Binary Collision Approximation Code for Simulating Ion Interactions with Rough Surfaces}



\author{
Jon Drobny,
Alyssa Hayes,
Davide Curreli,
David N. Ruzic
}

\address{
Nuclear, Plasma, and Radiological Engineering \\
University of Illinois at Urbana Champaign, Urbana, IL 61801
}

\begin{abstract}
Fractal TRIDYN (F-TRIDYN) is a modified version of the widely used Monte Carlo, Binary Collision Approximation code TRIDYN that includes an explicit model of surface roughness and additional output modes for coupling to plasma edge and material codes. Surface roughness plays an important role in ion irradiation processes such as sputtering; roughness can significantly increase the angle of maximum sputtering and change the maximum observed sputtering yield by a factor of 2 or more. The complete effect of surface roughness on sputtering and other ion irradiation phenomena is not completely understood. Many rough surfaces can be consistently and realistically modeled by fractals, using the fractal dimension and fractal length scale as the sole input parameters. F-TRIDYN includes a robust fractal surface algorithm that is more computationally efficient than those in previous fractal codes and which reproduces available experimental sputtering data from rough surfaces. Fractals provide a compelling path toward a complete and concise understanding of the effect that surface geometry plays on the behavior of plasma-facing materials. F-TRIDYN is a flexible code for simulating ion-solid interactions and coupling to plasma and material codes for multiscale modeling.
\end{abstract}


\end{frontmatter}



\section{Introduction}
\label{S:1}

Plasma-material interactions (PMI) are crucial to the operation of all plasma devices, including experimental fusion reactors. In fusion experiments, interactions with the the wall are responsible for drastic changes in plasma performance and confinement, including radiation losses via impurities. Energetic ions are primarily responsible for many phenomena important to fusion reactor performance, including sputtering and tungsten fuzz growth \cite{petty2015}. In addition to semi-empirical formulas for calculating sputtering yields such as the Yamamura formula \cite{yamamura1996}, the two main strategies for studying ion-solid interactions are Molecular Dynamics (MD) and Binary Collision Approximation (BCA) simulations \cite{robinson1994}.

While semi-empirical equations for sputtering yields can be used for simple systems, simulations allow one to study a significantly larger volume of the many-dimensional parameter space for ion-solid interactions. Simulations provide information, such as implantation depth profiles and Frenkel pair damage profiles, that cannot be captured in a single equation. MD simulations directly integrate the equations of motion for a number of interacting atoms. This direct integration, generally relying on ab initio or experimentally derived pair interaction potentials, is computationally difficult. MD simulation of timescales relevant to ion-solid interactions in the fusion community requires access to high-performance computers and, in some cases, prohibitively long computational times. For this reason, MD codes are not practical to use as directly coupled components of multiscale codes. BCA codes, on the other hand, treat the ion-solid interaction as a combination of elastic two-body ion-atom collisions and inelastic electronic stopping \cite{eckstein1991}. BCA codes are fast enough to run on-line with other codes and are statistically and physically accurate for a wide range of materials and ion energies \cite{moller1988}.

While SRIM, based on the TRIM code, is arguably the most popular BCA code in use today, it has been shown to be inaccurate, especially for sputtered atom angular distributions \cite{hofsass2014}. TRIDYN, and its successor SDTrimSP, do not share the errors present in SRIM. They are upgraded versions of TRIM to include a more accurate interaction potential and features such as dynamic surface composition \cite{moller1988}. Fractal TRIDYN (F-TRIDYN) is a new, further upgraded version of TRIDYN to include explicit fractal surface roughness and additional output modes. Fractal surface roughness was first implemented in a BCA code in FTRIM \cite{ruzic1989},\cite{ruzic1990}. FTRIM, in addition to being based on the older TRIM code, utilized a fractal surface algorithm that was of $O(N^2)$, where $N$ is the number of points that comprise the fractal surface, making it prohibitively expensive to use for large-scale contemporary simulations. The new surface algorithm implemented in F-TRIDYN, however, is of $O(N)$. In addition to the new fractal surface algorithm and the inherent advantages in accuracy of using TRIDYN as a base code, F-TRIDYN adds features that FTRIM does not have, including output lists of stopped projectile locations, stopped Primary Knock-on Atom (PKA) and Secondary Knock-on Atom (SKA) locations, Frenkel pair damage locations, and energy-angle distributions of the sputtered particles, all in three dimensions. These additional output features in particular make F-TRIDYN suitable for coupling to a wide variety of plasma and material codes as a wall boundary condition for energetic ions.

\section{Fractal Surface Roughness}

Fractals are mathematical objects that possess either self-similarity or self-affinity. Self-similarity is the property that elements of an object at different scales are exactly or approximately the same. Self-affinity is a weaker quality that allows for an object to scale differently along different spatial dimensions. Fractals can be characterized by a so-called fractal dimension. Fractal dimensions describe how the size of the space fractals occupy relates to the size of the space within which they are embedded. In particular, the fractal dimension of a curve describes how its total measured length changes depending on the length scale of the measurement. For example, smooth curves will be measured to have the same total length regardless of the length scale considered. Therefore, the space smooth curves occupy scales according to a power law of order equal to their topological dimension. Measurements of the length of fractal curves, however, will scale according to a power law of order greater than their topological dimension, meaning that their total length is not well-defined and depends on the observed length scale. In particular, they will scale according to a power law of order equal to their fractal dimension \cite{mandelbrot1967}. Intuitively, fractal dimension can be seen as a measure of the complexity or roughness of a curve. Determination of the dimension of simple geometric objects, such as points, lines, and planes is trivial. For fractals, however, there are multiple methods of measuring their dimensions, which are not, in general, equal. Throughout this paper, the fractal dimension referred to will be the box-counting dimension \cite{feder1988} as described hereafter.

\begin{figure}[ht] 
\includegraphics[width=\textwidth]{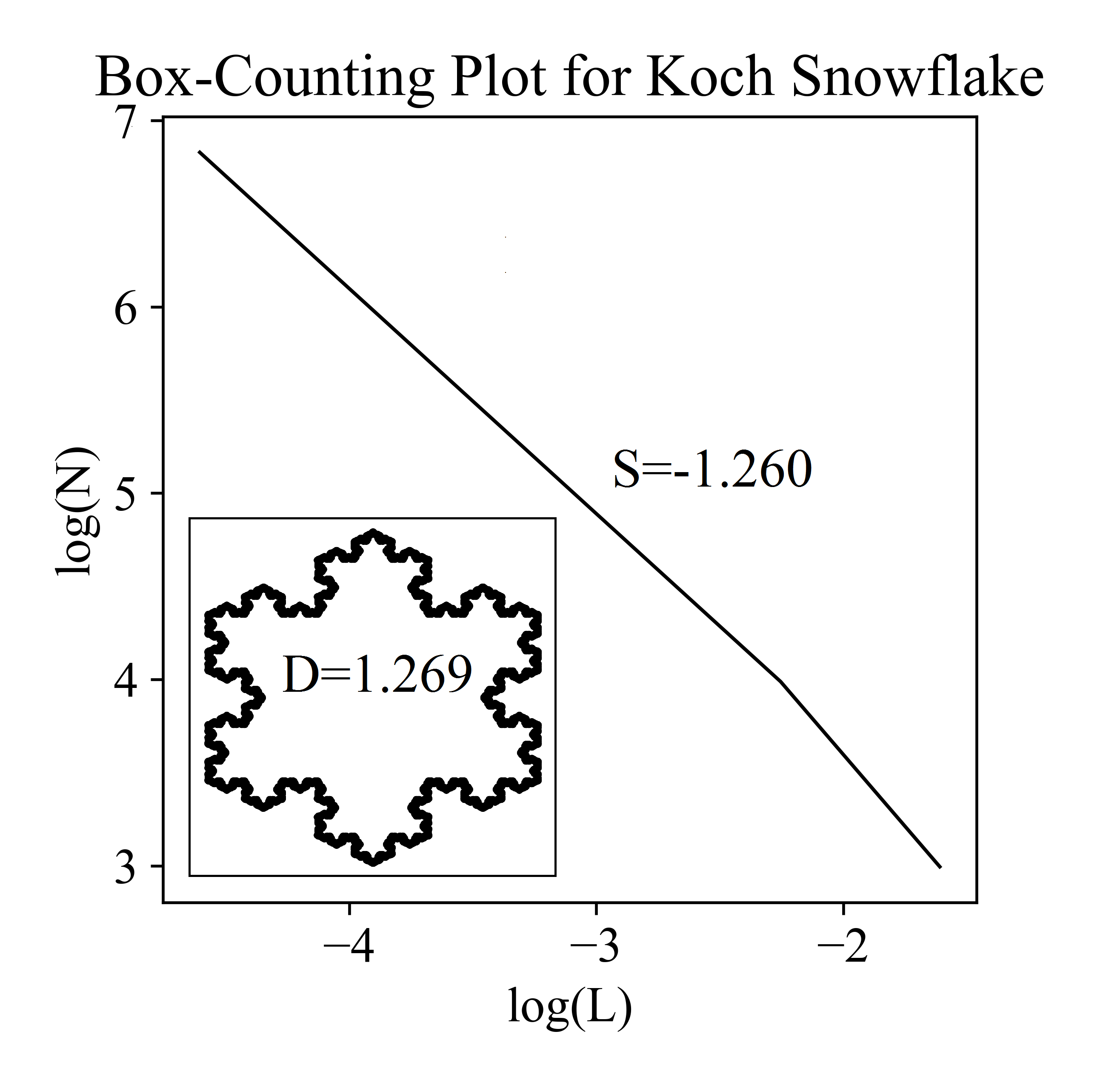}
\caption{The Koch snowflake (inset) and its corresponding box-counting plot. After extracting points from a rendered image of the Koch snowflake, the box-counting algorithm was used to measure the slope of the log-log plot of the number N of boxes of side length L occupied by the curve for a range of side lengths. The negative slope of the line, found via least squares linear regression, is equal to 1.260. The actual fractal dimension of the Koch snowflake is 1.269.}
\label{fig1}
\end{figure}

Calculation of the box-counting dimension is straightforward. For a given curve, separate the space it occupies into boxes of side length $L$. The number of boxes that are at all occupied by any portion of the fractal curve are counted as $N$. If one plots on a log-log scale the number of occupied boxes versus the box side length, the slope of any linear section of the resulting line will be equal to the negative fractal dimension in that region. Some curves may exhibit fractal behavior over a limited range of length scales. Outside of this range the box-counting plot will either have slope zero or be nonlinear \cite{mandelbrot1967}. The range of length scales for which a curve has a measurable, non-zero fractal dimension will be referred to as the fractal length scale. In Fig.~\ref{fig1}, the Koch Snowflake and its box-counting plot are shown.

	Many natural objects and physical phenomena are fractal at some length scale \cite{feder1988}. Particularly, fractals provide a consistent and realistic model for rough molecular surfaces \cite{avnir1984}, \cite{kajita2014}. For surfaces such as these, the roughness can be characterized by just two parameters, the fractal dimension and the fractal length scale. This allows complex geometry representative of the rough surface to be generated on the fly for any arbitrary fractal dimension. Additionally, fractal dimension is practically measurable in-situ. For most PMI experiments, it is not possible to measure surface morphology directly without breaking vacuum. Using the methods of Fractal BET adsorption theory \cite{pfeifer1989} \cite{hu2014}, the fractal dimension of a rough surface can be determined via the amount of adsorption onto the surface of differently sized gas molecules. This offers a compelling methodology to study the effect of rough surfaces on ion-solid interactions in a single machine by measuring surface roughness, performing ion irradiation, and performing subsequent surface diagnostics.

\section{Generation of Fractal Surfaces for F-TRIDYN}

	Fractal surfaces in F-TRIDYN are composed of two orthogonal cross-sections. Because a fractal surface will have an overall fractal dimension equal to the sum of the dimensions of each of its cross sections, two-dimensional fractal surfaces of dimension $2.0 \leq D_{2D} \leq 3.0$ can be represented by a combination of a dimension 1.0 cross-section and a dimension $D_{2D}-1.0$ cross section. For this reason, fractal surfaces in F-TRIDYN are represented using a single fractal cross section, spanning the $Y-Z$ plane. As long as the fractal dimension of a surface is homogeneous across the target, this is a reasonable approximation for a Monte Carlo code such as F-TRIDYN. Although surface roughness in F-TRIDYN is two-dimensional, the addition of complete spatial particle tracking in F-TRIDYN over TRIDYN allow for the possibility of three dimensional surface implementation in the future.
    
    Fractal cross sections are created for F-TRIDYN using the fractal generator method popularized by Mandelbrot \cite{mandelbrot1967}. This method iteratively combines an initial piecewise curve, known as the generator, to construct a fractal curve. A true fractal would require iteration ad infinitum, but for the description of atomic surfaces the generator need only be iterated until the distance between points is approximately equal to desired minimum interatomic distance. A recursive algorithm is used to construct the curve as input to the code. The initial curve is translated, rotated, and scaled to the local angle and segment length for each line segment of the previous generation.
    
\begin{figure}[ht] 
\includegraphics[width=\textwidth]{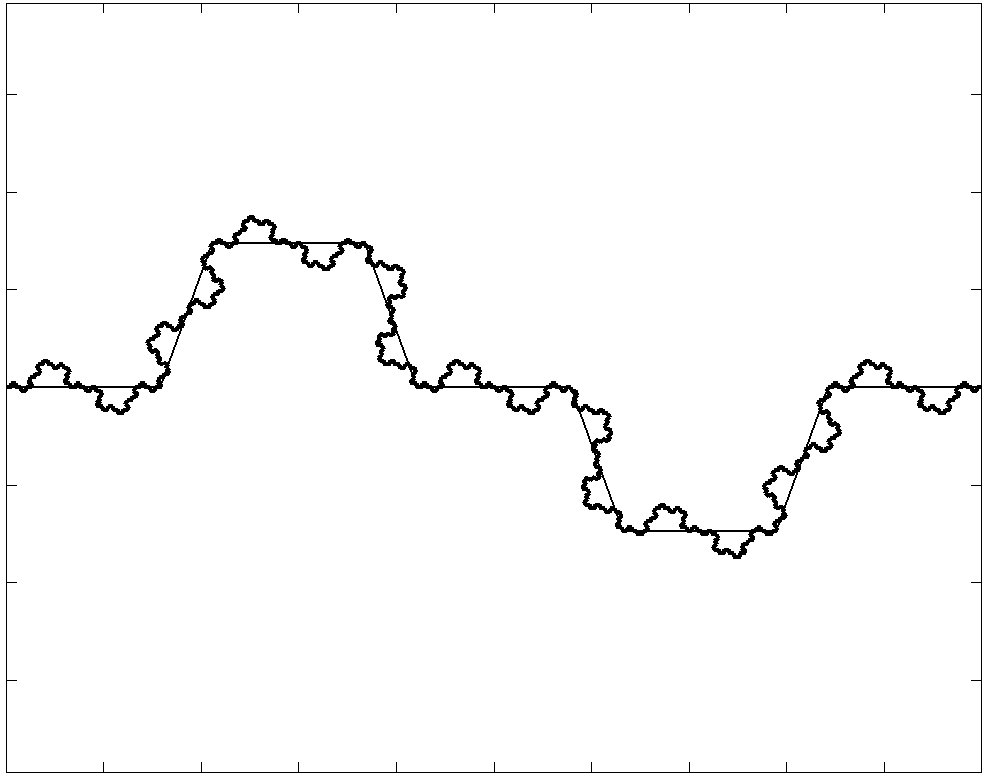}
\caption{An example of the default fractal generator in F-TRIDYN and its subsequent iterations with fractal dimension $D=1.18$. This fractal generator is used because it is qualitatively similar to molecular surfaces and because it can span, without self-intersections, the widest range of fractal dimensions (up to approximately $D=1.8$) of all tested fractal generators.}
\label{fig2}
\end{figure}

\section{Implementation of Fractal Surface Roughness in F-TRIDYN}

	Rough surfaces are included in F-TRIDYN via a parallelized implementation of the even-odd algorithm developed for solving the point-in-polygon problem in computer graphics \cite{sutherland1974}, \cite{vandewettering1990}. To determine whether a given particle is inside or outside the surface, a ray is cast upwards from the particle’s position and the number of intersections it makes with the surface is determined. A simple fact, provable from the Jordan Curve Theorem \cite{vandewettering1990}, is that if the number of intersections is odd, the particle’s position lies inside the surface, and if it is even, it lies outside the surface. This algorithm is O(N), where N is the number of points in the piecewise curve. F-TRIDYN can alternatively use any arbitrary piecewise curve for the rough surface. Surfaces for this work were created exclusively using fractal rough surfaces created via the fractal generation algorithm.
    
	Solid lattices in TRIDYN are treated as amorphous, and the locations of collision partners are chosen randomly from the number density of the local compositional layer \cite{moller1988}. If a collision partner is created above the surface in TRIDYN, determined by a simple inequality, the impact parameter for the collision is set to an arbitrarily high value such that the effect of the collision is negligible. The same method is used in F-TRIDYN, but the determination of whether the particle is inside or outside of the surface is performed using the methodology outlined above.

\section{Results for Sputtering Yields}

\begin{figure}[ht] 
\begin{tabular}{ccc}
\includegraphics[width=0.33\textwidth]{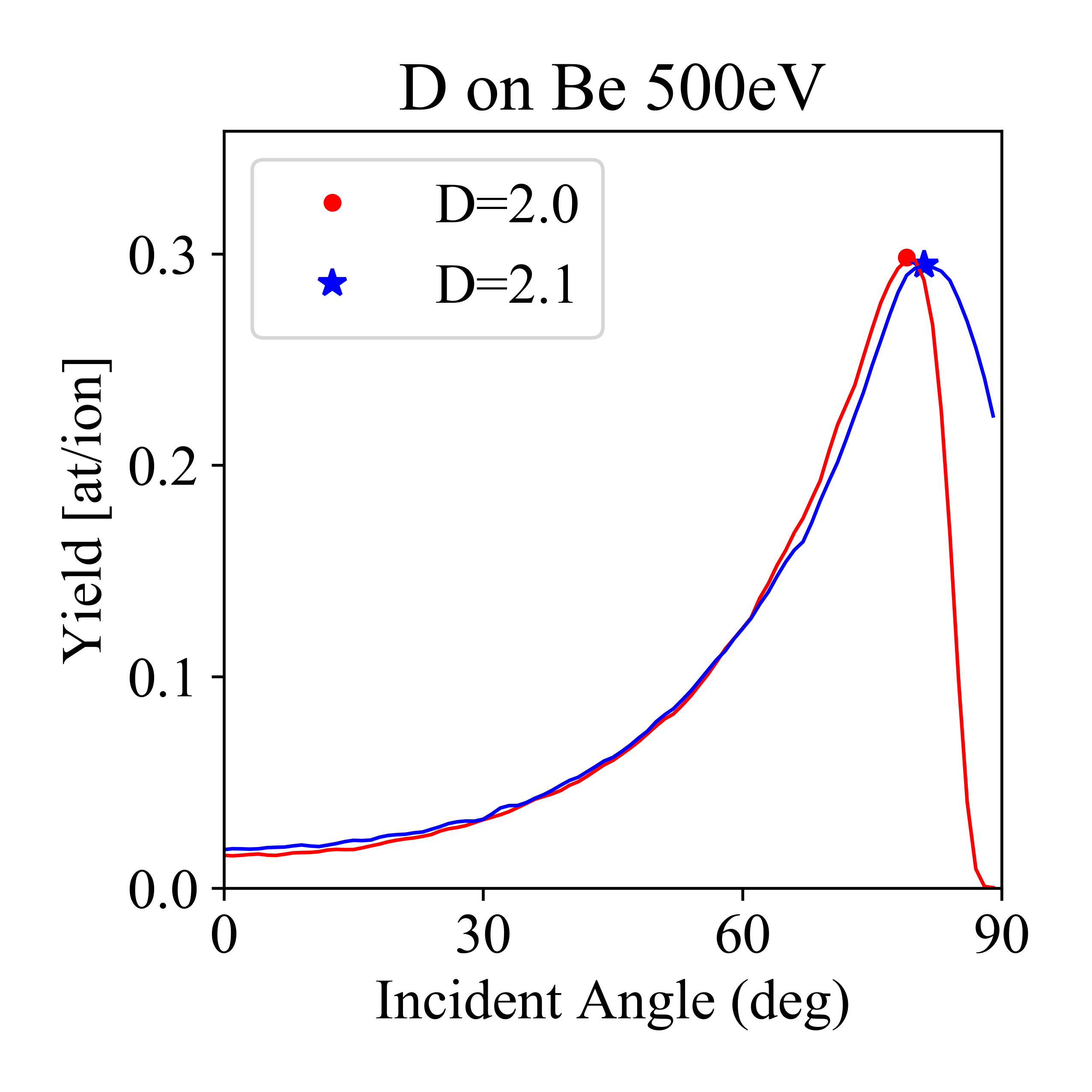} &
\includegraphics[width=0.33\textwidth]{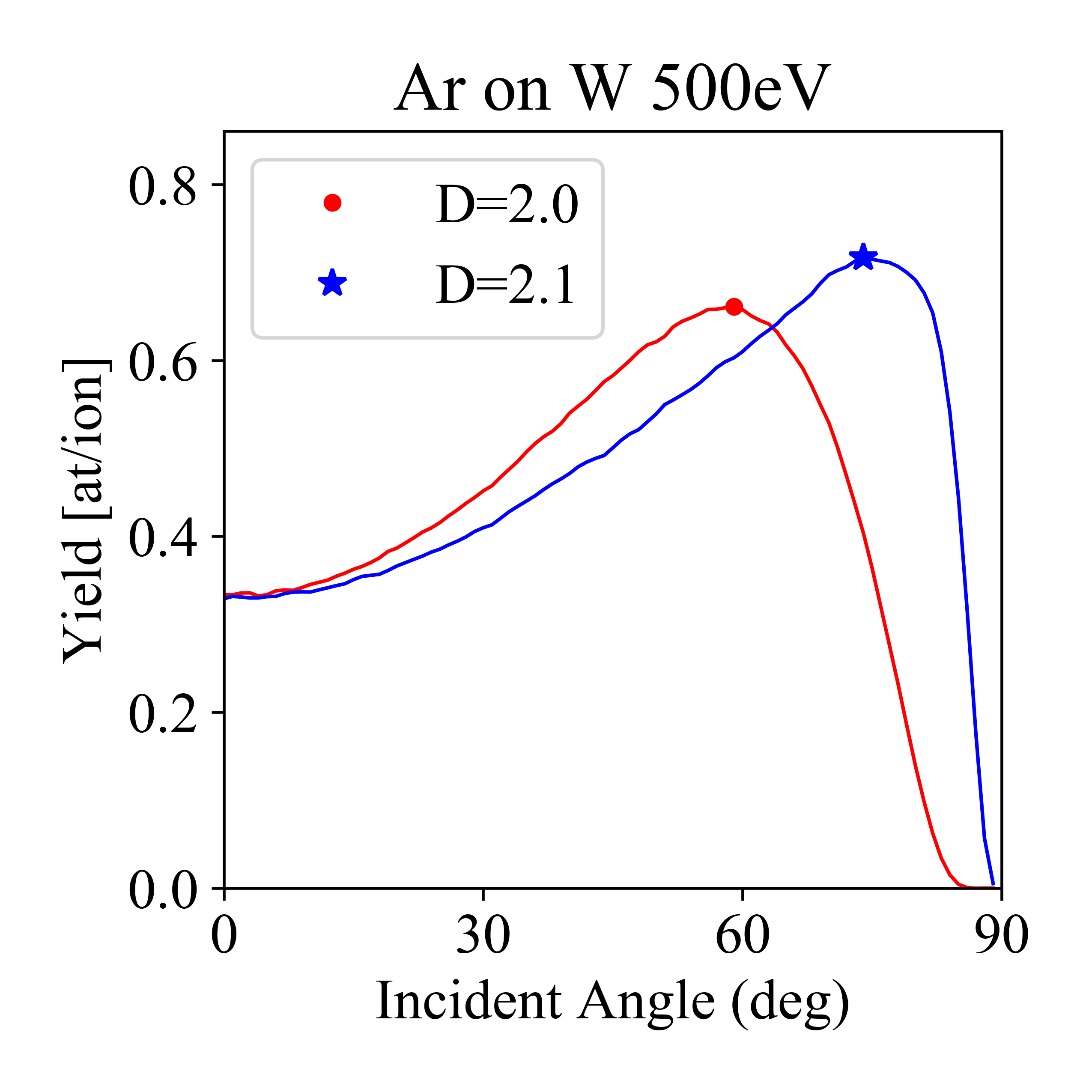} &
\includegraphics[width=0.33\textwidth]{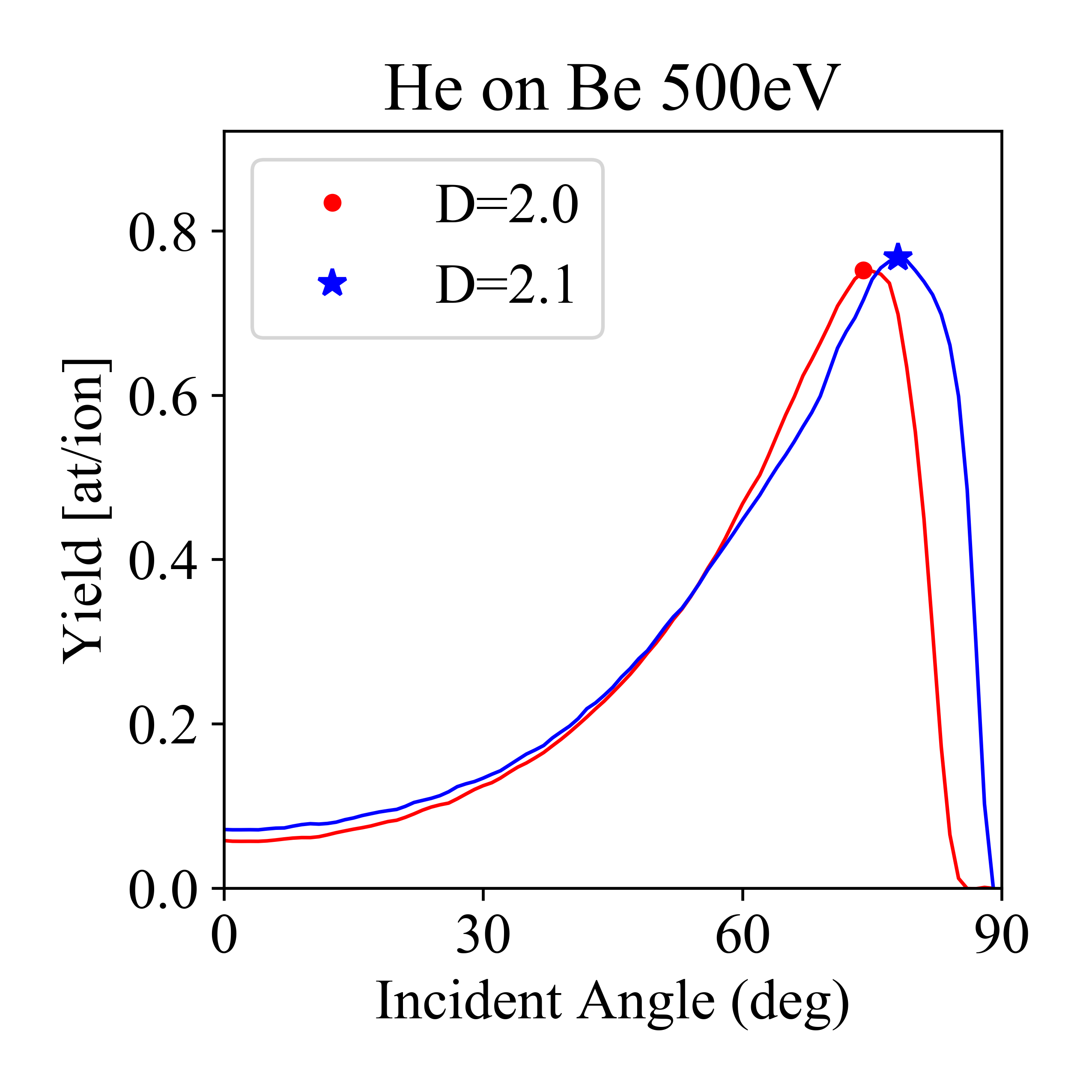} \\
(a) & (b) & (c)
\end{tabular}
\caption{Results of the sputtering yield as a function of angle of incidence for: (a) Deuterium on Beryllium, (b) Argon on Tungsten, and (c) Helium on Beryllium at 500 eV. F-TRIDYN results reproduce two key features seen in other works \cite{wei2009}, \cite{kustner1999}, \cite{makeev2004}, namely: reduction in yield for some systems at medium angle of incidence and the shift of the maximum sputtering yield to the right.}
\label{fig3}
\end{figure}

\begin{figure}[ht] 
\includegraphics[width=\textwidth]{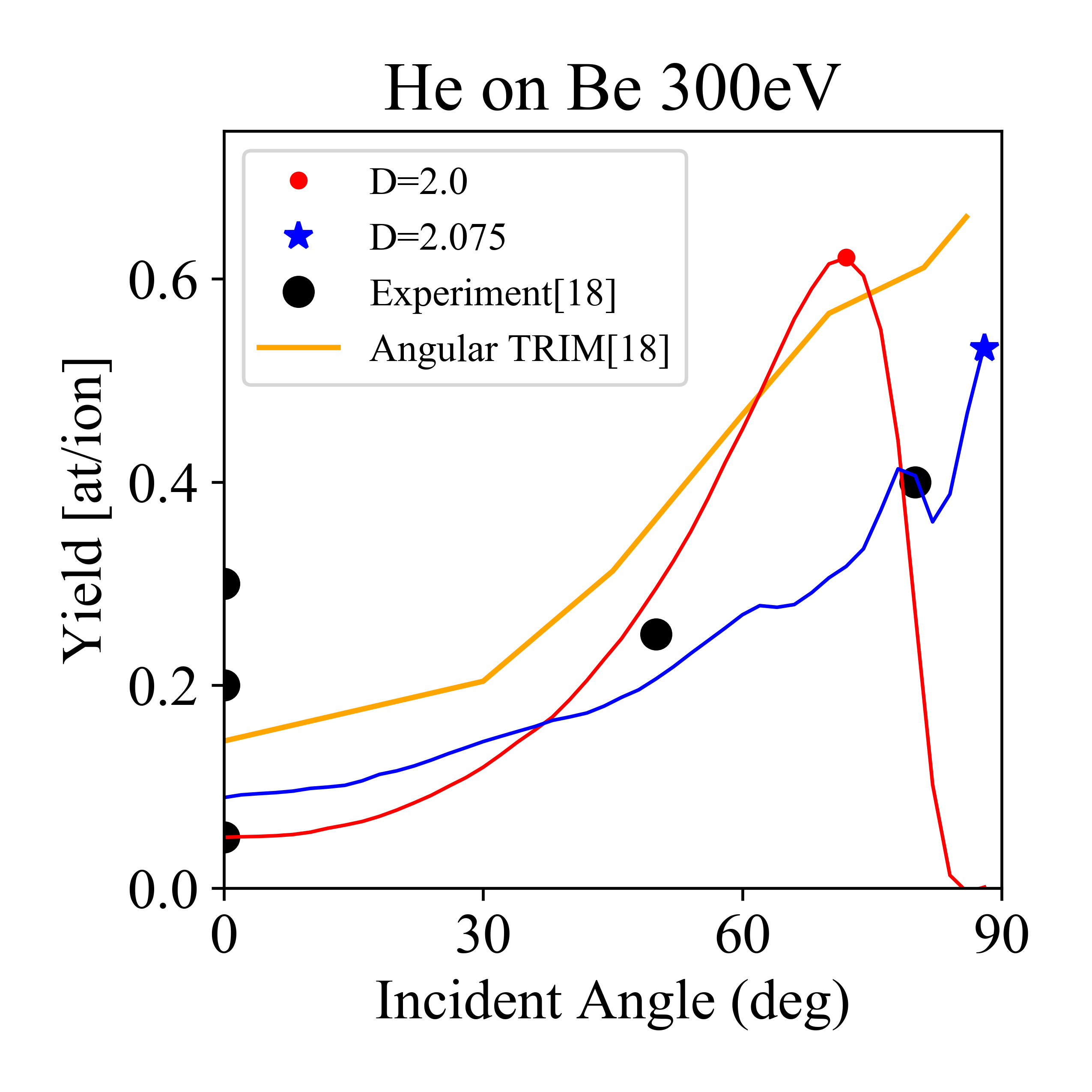}
\caption{Comparison of sputtering yield as a function of angle of incidence for: (1) F-TRIDYN for a smooth surface (``$D = 2.0$'', red curve), (2) F-TRIDYN for a rough surface (``$D = 2.075$'', blue curve), (3) experimental data \cite{kustner1999} (``Experiment\cite{kustner1999}'', black dots), and (4) the semi-empirical local angle-of-incidence TRIM simulation done by Küstner et al. (``Angular TRIM \cite{kustner1999}'', yellow curve).  F-TRIDYN replicates the behavior seen in both the experiment and the alternate simulation method, although validation is weak due to the small number of experimental data points.}
\label{fig4}
\end{figure}

Figures \ref{fig3} (a)-(c) compare the angular sputtering yield for two surface roughnesses of three ion-target systems: Helium on Beryllium, Deuterium on Beryllium, and Argon on Tungsten respectively. As observed in other simulations and experiments, the peak of maximum sputtering yield shifts to the right \cite{wei2009}, \cite{kustner1999}, \cite{makeev2004}. In Fig.~\ref{fig4}, F-TRIDYN results are compared to an alternate simulation method by Küstner et al., namely using STM (Scanning Tunneling Microscope) images to reconstruct local angle of incidence distributions for the surface and using those distributions as inputs for TRIM, and an experiment. Sputtering yields are difficult to measure and prone to significant experimental error. Nevertheless, F-TRIDYN compares quantitatively favorably to Küstner et al.'s simulation technique and qualitatively recreates the trends seen in the experiment. The fractal dimension for this comparison was chosen heuristically based on the standard deviation of heights, as the fractal dimension in that study was not measured, only the deviation in surface heights. This leads to the possibility of further increases in accuracy if the fractal dimension and length scale are measured directly. In contrast to Küstner et al.'s simulation, F-TRIDYN allows one to characterize a surface with two parameters, and allows for one to simulate rough surfaces even when their exact morphology is unknown. 

\begin{figure}[ht] 
\begin{tabular}{ccc}
\includegraphics[width=0.33\textwidth]{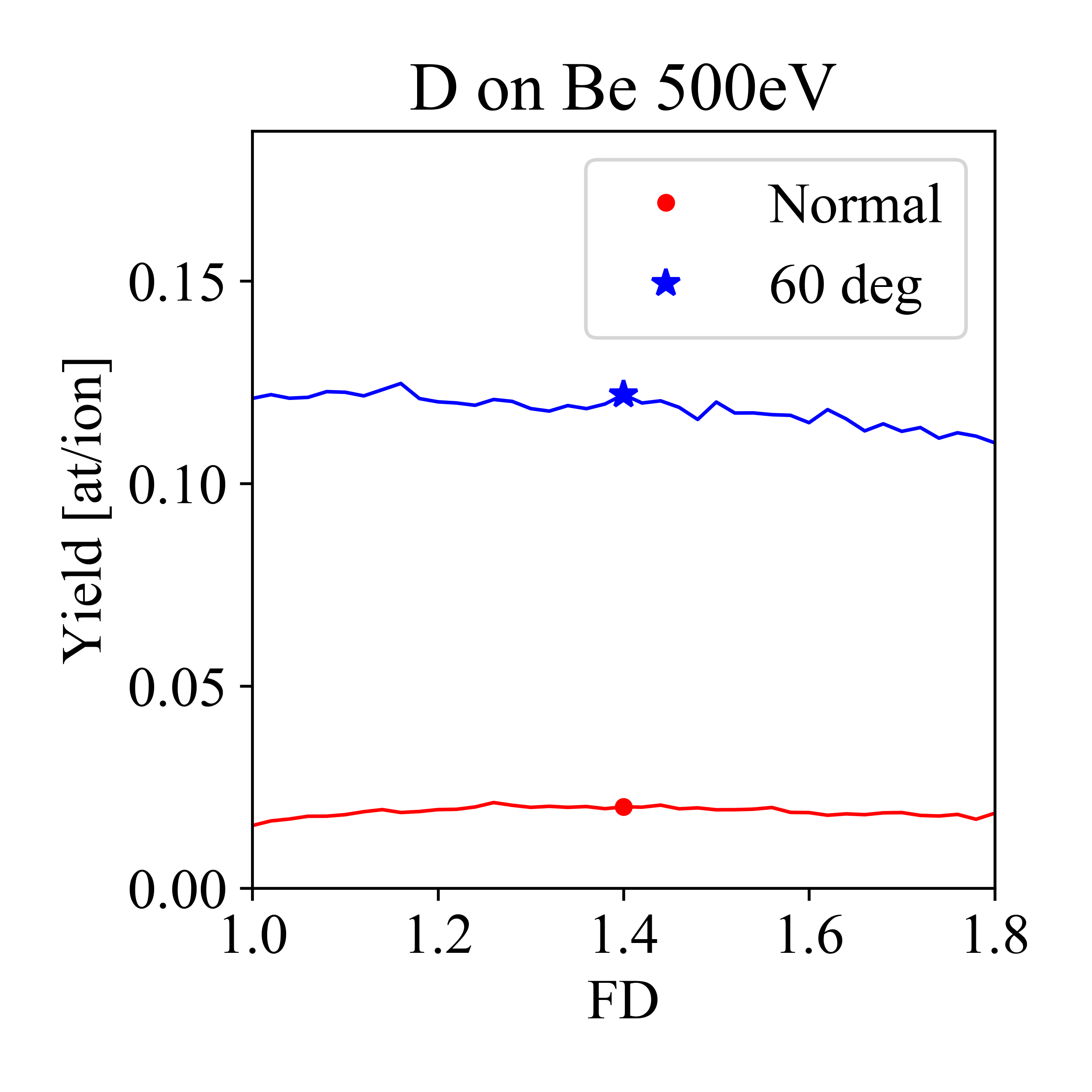} &
\includegraphics[width=0.33\textwidth]{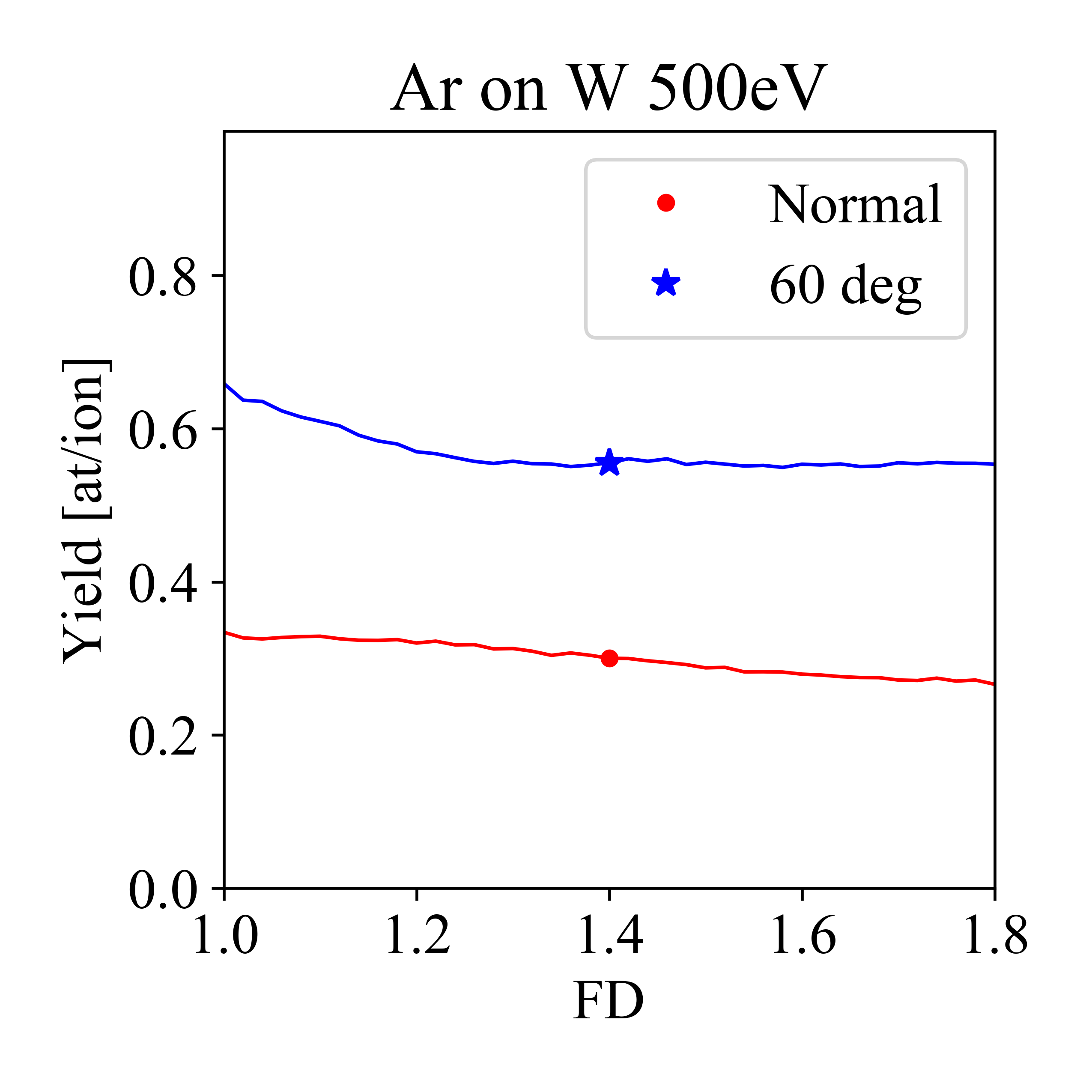} &
\includegraphics[width=0.33\textwidth]{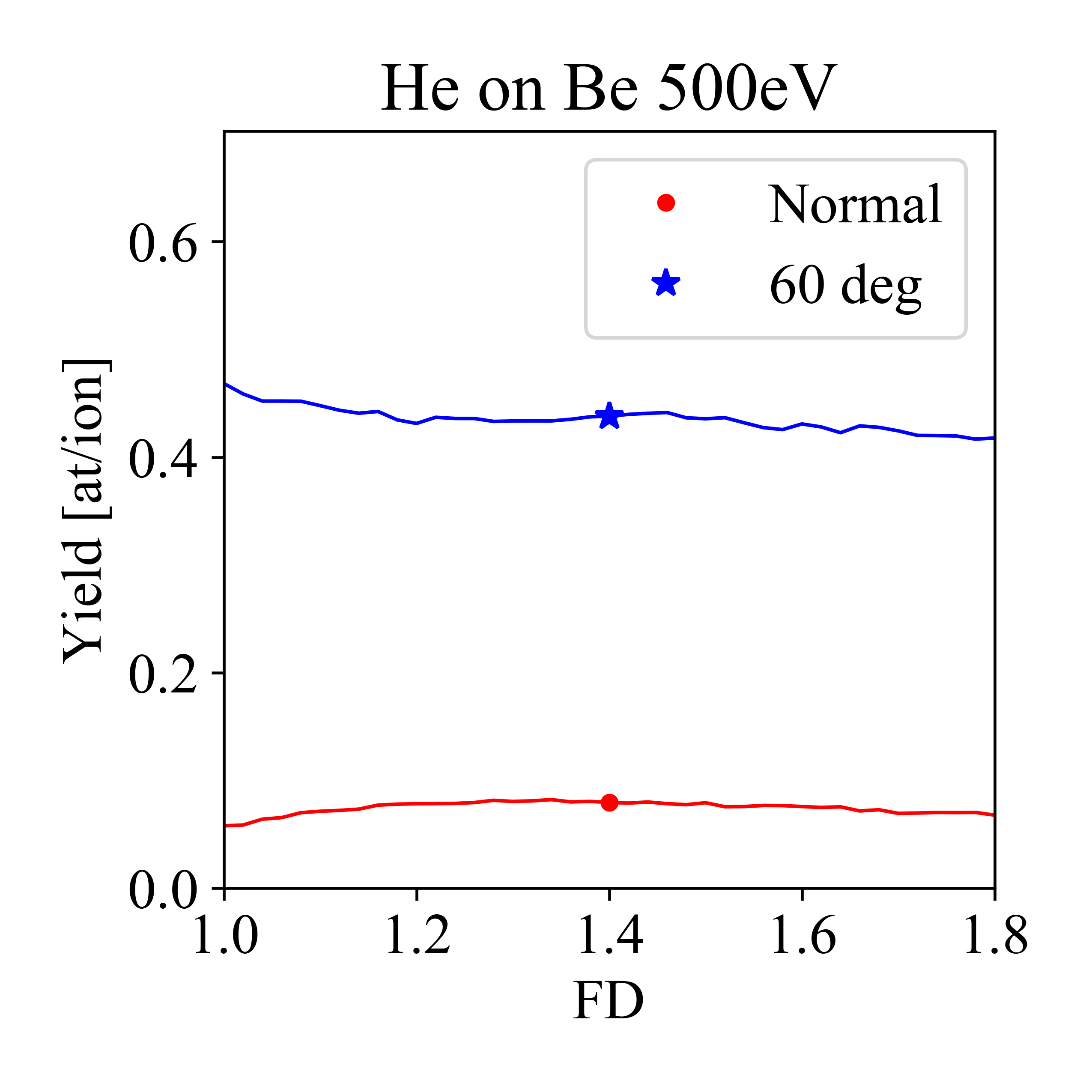} \\
(a) & (b) & (c)
\end{tabular}
\caption{Results of the sputtering yield as a function of fractal dimension for: (a) Deuterium on Beryllium, (b) Argon on Tungsten, and (c) Helium on Beryllium at 500 eV. F-TRIDYN results qualitatively match those of FTRIM [7], but experimental studies that include a measurement of fractal dimension are necessary to validate the choice of fractal length scale used in these simulations (40 Angstroms).}
\label{fig5}
\end{figure}

\begin{figure}[ht] 
\includegraphics[width=\textwidth]{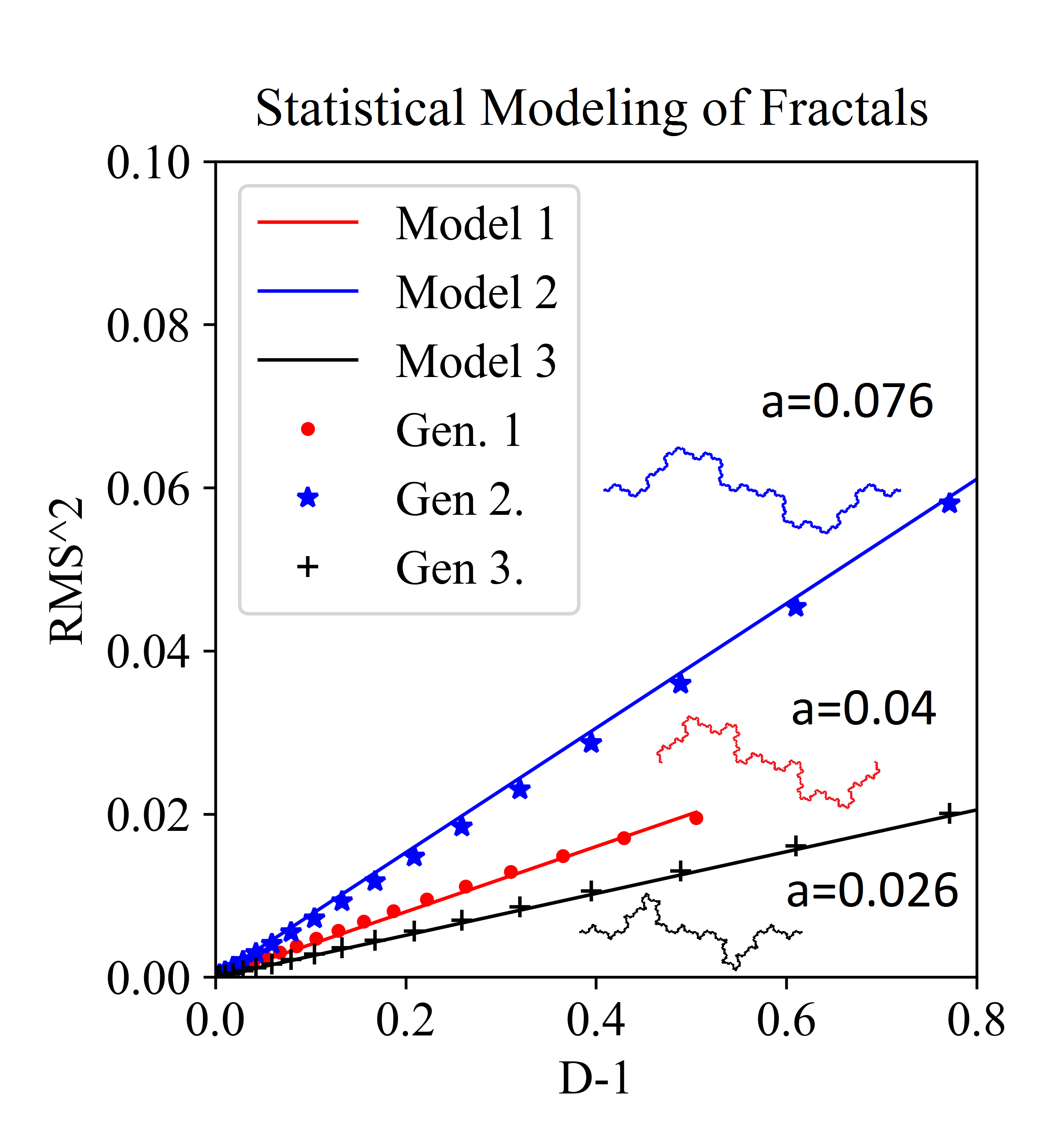}
\caption{A comparison of the constant $a$ for estimating the standard deviation of heights of 3 different fractal generators. F-TRIDYN’s default fractal generator is represented by the uppermost curve with slope 0.076.}
\label{fig6}
\end{figure}

Fractal dimension has a nonlinear relationship to other measures of roughness, such as the RMS roughness. In particular, the standard deviation of heights of the points of a fractal surface created using the fractal generator method is proportional to the square root of the fractal dimension. The relationship between the standard deviation of heights squared and the fractal dimension is plotted for three different fractal generators in Fig.~\ref{fig6}. The constant $a$ depends only on the shape of the fractal generator and the fractal length scale. With the limit of $1.0 \leq D \leq 2.0$ for a fractal curve, this means that even small fractal dimensions can play a significant role in ion-solid interactions. For the same three systems as before, the effect of fractal dimension on sputtering yield for two angles of incidence is plotted in ~\ref{fig5}. With a fractal length scale of 40 Angstroms, the result for high angles of incidence is overall a slight reduction in yield.

\begin{figure}[ht]
\begin{center}
\begin{tabular}{ccc}
\includegraphics[width=0.33\textwidth]{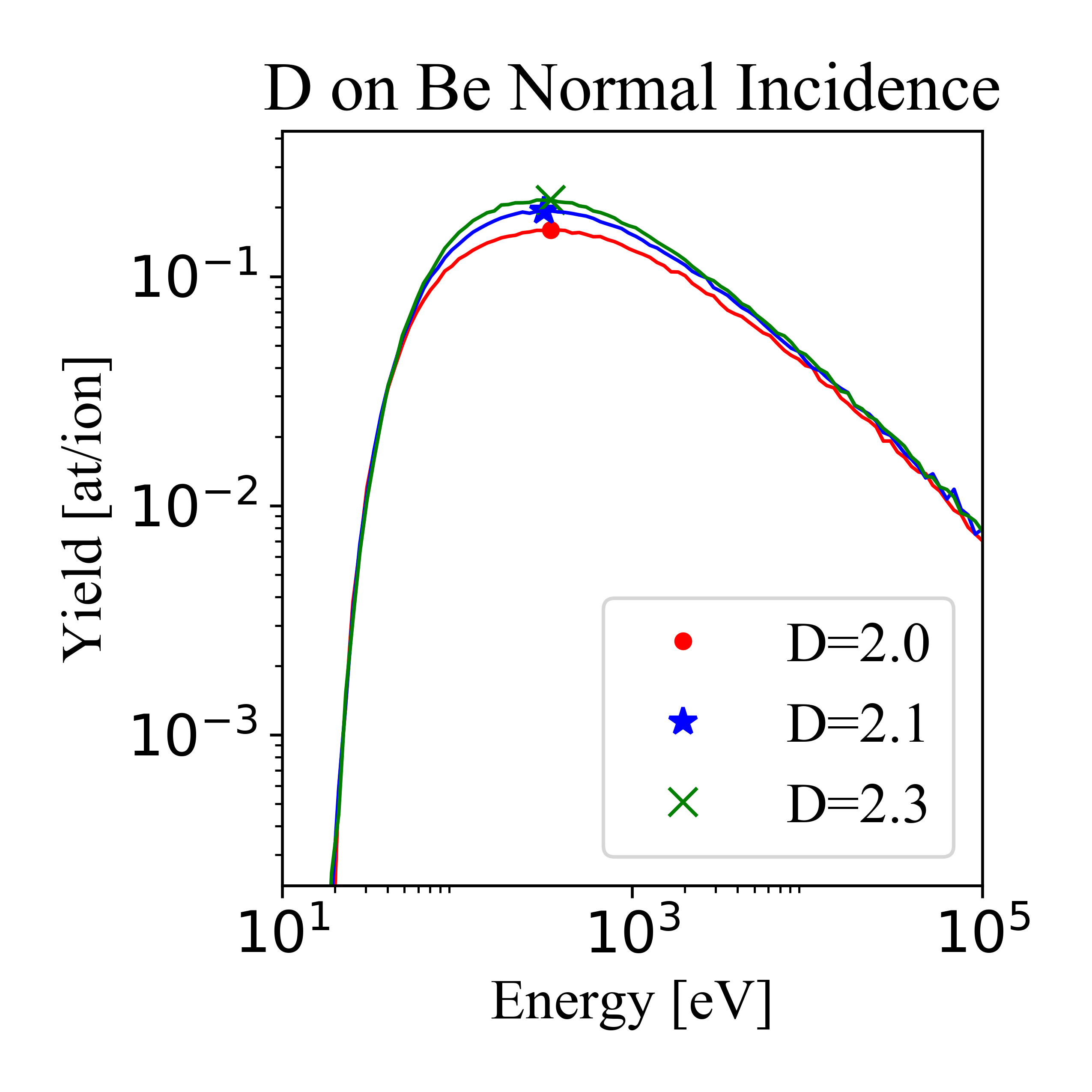} &
\includegraphics[width=0.33\textwidth]{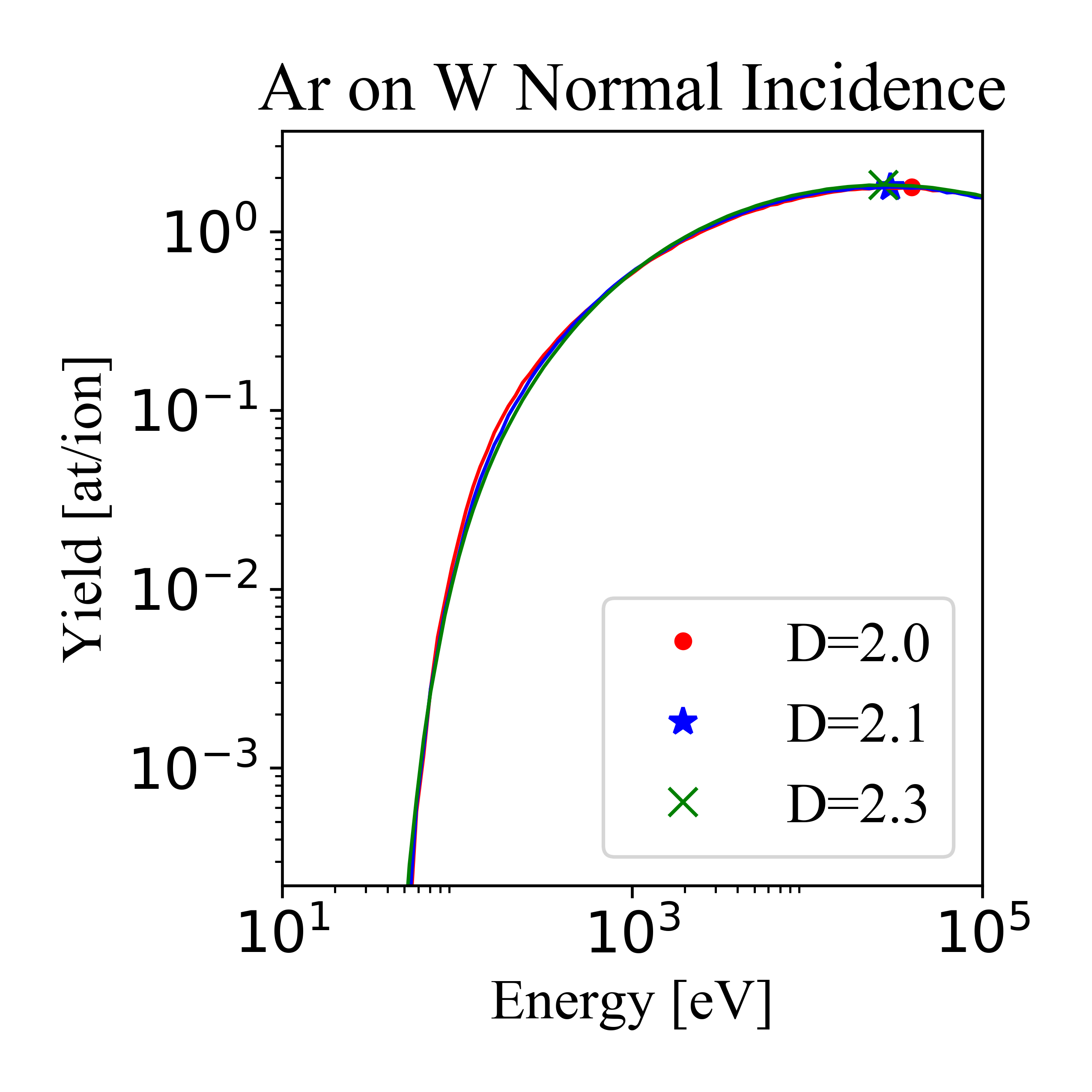} &
\includegraphics[width=0.33\textwidth]{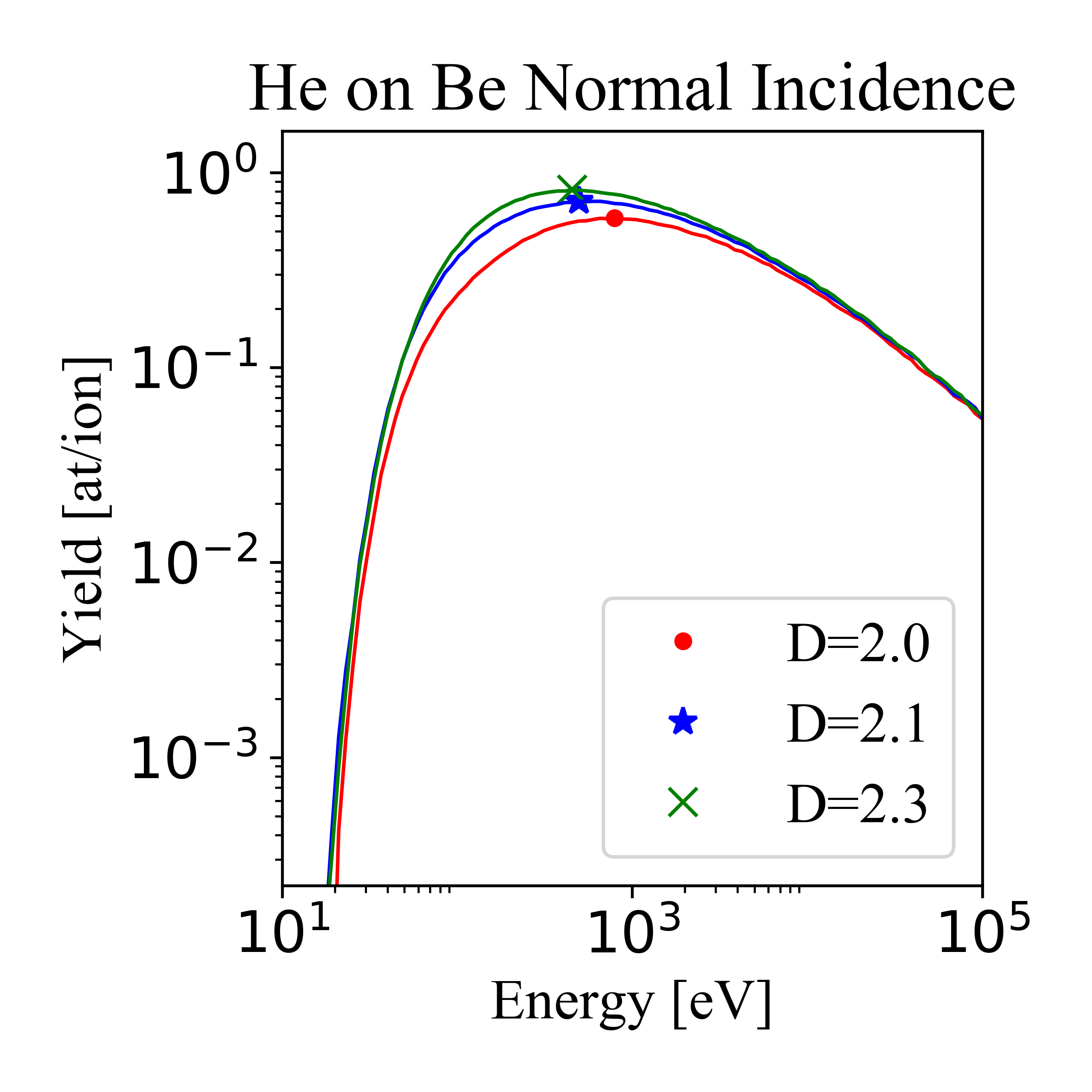} \\
(a) & (b) & (c)
\end{tabular}
\caption{Results of the sputtering yield as a function of incident ion energy for: (a) Deuterium on Beryllium, (b) Argon on Tungsten, and (c) Helium on Beryllium.}
\label{fig7}
\end{center}
\end{figure}

Figure \ref{fig7} (a)-(c) compare the sputtering yield versus energy for three surface roughnesses at normal incidence for the same systems as presented above. For most energies and systems, surface roughness will decrease the sputtering yield, as there is a greater chance that atoms emitted from the surface will be recaptured at a different location. For Helium and Deuterium on Beryllium, the sputtering yield increased with roughness for a large range of energies. Argon on Tungsten sputtering yields remained relatively constant with changes in surface roughness. Due to the speed of the fractal surface algorithm, reasonable simulation statistics to capture sputtering yields near threshold, an important energy regime for future fusion reactors, can be accomplished with F-TRIDYN.

\section{Conclusions}

F-TRIDYN is a flexible BCA code for the simulation of ion-solid interactions with a rough surface. It performs within experimental error for systems where there is available sputtering yield data that includes an explicit measurement of surface roughness. In order to further validate the fractal surface roughness model, however, significantly more experimental data is needed. Surface roughness is predicted to play a significant role in the performance of future fusion reactors as energetic ions drive significant morphological evolution in Plasma Facing Components (PFCs). F-TRIDYN is designed for simple coupling to other codes, adding 3D spatial tracking and output of projectiles, distributions of sputtered atoms, PKAs, and SKAs in addition to Frenkel Pair damage. As a component in a multiscale code, F-TRIDYN will be useful for simulation of the atomic timescale effects of energetic ions on a material, such as erosion, implantation, and backscattering, and will act as the boundary between plasma edge and material bulk codes. 

\section*{Acknowledgments}
This work was supported by the PSI-SciDAC Project funded by the U.S. Department of Energy through contract DOE-DE-SC0008658.





\bibliographystyle{model1-num-names}
\bibliography{library.bib}







\end{document}